\documentclass[a4paper,conference]{IEEEtran}
\usepackage[T1]{fontenc}
\usepackage[utf8]{inputenc}
\usepackage{amsmath}
\usepackage[super]{nth}
\usepackage{paralist}
\usepackage{pgfplots}
\usepackage{tikz}
\usetikzlibrary{calc}
\usetikzlibrary{positioning}
\usetikzlibrary{shapes}
\usetikzlibrary{snakes}

\title{Interoperator fixed-mobile network sharing}

\author{
  \IEEEauthorblockN{Ireneusz Szcześniak, Piotr Chołda, Andrzej R.~Pach}
  \IEEEauthorblockA{AGH University of Science and Technology\\
    Department of Telecommunications\\
    al.~Mickiewicza 30\\
    30-059 Krakow\\
    Poland}
  \and
  \IEEEauthorblockN{Bożena Woźna-Szcześniak}
  \IEEEauthorblockA{Institute of Mathematics and Computer Science\\
    Jan Długosz University\\
    al.~Armii Krajowej 13/15\\
    42-200 Częstochowa\\
    Poland}
}

\begin{document}

\tikzstyle{circ} = [shape = circle, draw, inner sep = 1 pt]
\tikzstyle{circ1} = [circ]
\tikzstyle{circ2} = [circ, fill = black!30]
\tikzstyle{icwave} = [->, > = stealth, red, decorate,
  decoration = {snake, segment length = 2 mm, post length = 1 mm}]
\tikzstyle{muwave} = [<->, > = stealth, gray, decorate,
  decoration = {snake, amplitude = 0.25 mm, segment length = 1 mm,
    post length = 1 mm, pre length = 1 mm}]
\tikzstyle{iclink} = [->, > = stealth, red]
\tikzstyle{mycloud} = [cloud, cloud puffs = 15, cloud ignores
  aspect, minimum width = 3 cm, minimum height = 1.25 cm, align = center,
  draw]
\tikzstyle{switch} = [draw, inner sep = 0, minimum width = 0.5 cm,
  minimum height = 0.5 cm]
\tikzstyle{router} = [draw, circle, inner sep = 0, minimum size = 0.5 cm]
\tikzstyle{fmn} = [draw, ellipse, align = center, minimum width = 1.5 cm]
\tikzstyle{arn} = [circle, draw, inner sep = 1.5 pt]
\tikzstyle{prn} = [arn, fill = black]

\newcommand{\BS}[1]{%
  \begin{tikzpicture}[#1]
    \fill (0, 0) circle (5 mm);
    \draw ([shift = (-45:15 mm)] 0, 0) arc (-45:45:15 mm);
    \draw ([shift = (135:15 mm)] 0, 0) arc (135:225:15 mm);
    \draw ([shift = (-45:10 mm)] 0, 0) arc (-45:45:10 mm);
    \draw ([shift = (135:10 mm)] 0, 0) arc (135:225:10 mm);
    \coordinate (ll) at ($ (0, 0) + (-15 mm, -40 mm) $);
    \coordinate (rl) at ($ (0, 0) + (15 mm, -40 mm) $);
    \draw (0, 0) -- (ll) -- ($ (0, 0)!0.5!(rl) $) -- ($ (0, 0)!0.25!(ll) $);
    \draw (0, 0) -- (rl) -- ($ (0, 0)!0.5!(ll) $) -- ($ (0, 0)!0.25!(rl) $);
  \end{tikzpicture}%
}

\newcommand{\FU}[1]{%
  \begin{tikzpicture}[#1]
    \draw (-20 mm, -15 mm) rectangle (20 mm, 15 mm);
    \draw (-15 mm, -5 mm) rectangle (0 mm, 10 mm);
    \draw (-25 mm, 13 mm) -- (0, 30 mm) -- (25 mm, 13 mm);
    \draw (5 mm, -15 mm) -- (5 mm, 10 mm) -- (15 mm, 10 mm) -- (15 mm, -15 mm);
  \end{tikzpicture}%
}

\newcommand{\MU}[1]{%
  \begin{tikzpicture}[#1]
    \draw (3 mm, 0) arc (0:90:3 mm);
    \draw (5 mm, 0) arc (0:90:5 mm);
    \draw (-10 mm, -15 mm) rectangle (0 mm, 0 mm);
    \draw (-8 mm, -13 mm) rectangle (-2 mm, -2 mm);
  \end{tikzpicture}%
}

\newcommand{\VDOTS}[1]{%
  \begin{tikzpicture}[#1]
    \fill [black] ($ (0, 0) + (0, 4 pt) $) circle (1 pt);
    \fill [black] (0, 0) circle (1 pt);
    \fill [black] ($ (0, 0) + (0, -4 pt) $) circle (1 pt);
  \end{tikzpicture}%
}

\maketitle

\begin{abstract}
We propose the novel idea of interoperator fixed-mobile network
sharing, which can be software-defined and readily-deployed.  We study
the benefits which the sharing brings in terms of resiliency, and show
that, with the appropriate placement of a few active nodes, the mean
service downtime can be reduced more than threefold by providing
interoperator communication to as little as one optical network unit
in one hundred.  The implementation of the proposed idea can be
carried out in stages when needed (the pay-as-you-grow deployment),
and in those parts of the network where high service availability is
needed most, e.g., in a business district.  While the performance
should expectedly increase, we show the resiliency is brought almost
out of thin air by using redundant resources of different operators.
We evaluated the service availability for 87400 networks with the
relative standard error of the sample mean below 1\%.
\end{abstract}



\section{Introduction}


A fixed-mobile network (FMN) delivers services for fixed users (FUs)
to their premises, and for mobile users (MUs) to their user equipment
(UE) such as mobile phones or mobile routers.  This fixed-mobile
convergence allows a network operator to consolidate and simplify
business.  A FMN is composed of the radio access network (RAN), e.g.,
the Long-Term Evolution (LTE) network, and the backhaul, e.g., a
passive optical network (PON).  The backhaul connects both FUs and the
RAN base stations (BSs).


Maintaining and upgrading a FMN is expensive, and the scarce radio
spectrum for RANs is getting more expensive.  For these reasons, the
interoperator sharing of the network and spectrum is gaining
prominence, because it offers to lower costs and increase revenue
through better network and spectrum utilization
\cite{10.1109/JPROC.2014.2302743}.  But sharing can have various
forms.


Currently network sharing between competing operators is a fact
\cite{10.1109/MCOM.2011.6035827}, but it is limited to the physical
network infrastructure only (buildings, towers, etc.).  Some operators
\emph{merge} their networks into a single network, and then own and
use it together in a marriage-like fashion.  These forms of sharing
are of the legal, not technological nature.


We concentrate on the sharing enabled by technology, where an operator
is able to temporarily rent resources from other operators.  Unlike in
network merging, an operator can use a resource without owning it.  A
resource can be anything used to implement a service: it can be a
fixed or mobile resource, including spectrum.  The traffic and
technical difficulties which operators experience at a given time and
place can differ substantially indeed between different operators, and
the interoperator sharing would allow the operators to do better.


Sharing in FMNs pertains to RAN sharing and backhaul sharing.  In
RANs, the dynamic spectrum access (DSA) allows for various forms of
spectrum sharing.  Sharing of the backhaul could be realized with a
virtual local area network (VLAN) or a Carrier Ethernet network.
Software-definition augments the implementation.


The resiliency of the future FMN is crucial, but in the currently
deployed FMNs it is missing.  For instance, the currently deployed LTE
is not resilient, and so are not the PONs.  Resiliency is one of the
key requirements of the fifth generation (5G) networks
\cite{10.1109/MNET.2012.6172271}, and of the next generation PONs
(NG-PONs) \cite{10.1109/MCOM.2009.5307465}.


FMNs are being broadly researched and developed to deliver the
required performance and resiliency \cite{10.1109/JPROC.2012.2185769}.
Radio access technologies (RATs) have been proposed to use
cognitivity, virtualization, coordinated multipoint transmission
(CoMT), and more sophisticated modulation formats.  The backhaul is
evolving from the copper or microwave networks to passive optical
networks (PONs), and even possibly to radio-over-fiber (RoF) networks
\cite{10.1109/SURV.2013.013013.00135}.  PONs are currently being
deployed as the backhaul, and the NG-PONs are being intensively
researched for FMNs \cite{10.1109/JLT.2010.2050861}.  To the best of
our knowledge, no interoperator FMN sharing has been proposed before.


\emph{Our contribution is the novel idea of interoperator FMN sharing,
and the evaluation of the benefits the sharing brings in terms of
resiliency}.  The benefits are mainly realized by the communication
between different operators either wirelessly or optically.  Beside
the expected gain in performance, we argue that the resiliency is
almost there: the currently-deployed FMNs of different operators have
evolved independently and redundantly, and when they are shared, their
redundancy can be used to implement resiliency.


The article is organized as follows.  First, in the following Section
\ref{related} we review key related works, and in Section
\ref{sharing} we describe the proposed interoperator FMN sharing.
Next, in Section \ref{scenarios} we describe the evaluation setting,
and in Section \ref{results} we report on the obtained numerical
results.  Finally, Section \ref{conclusion} concludes the article.


\section{Related works}
\label{related}


Mobile network sharing has long been used, allowing for roaming or
virtual mobile network operators to exist, where a mobile operator
accepts traffic directly from the users of a different operator.  In
\cite{10.1109/MCOM.2011.6035827} the authors study the virtualization
support for this traditional sharing.  In \cite{6533005} the authors
discuss novel FMN architectures.  The hallmark of our proposed sharing
is the \emph{interoperator communication}, where traffic is exchanged
by different operators between their access networks.


In \cite{10.1109/TR.2011.2134210}, the authors propose a number of
wireless protection methods for FMNs.  There a single network is
considered, without sharing it with a different operator.  Wireless
access points connected to a PON are allowed to offer backup
connectivity to those wireless access points which lost the PON
connectivity.  These methods do not protect against, for example, the
failure of the feeder fiber, while our method does.


The various forms of DSA have been widely embraced by researchers,
industry and legislatures, and are regarded as the key enablers of 5G.
DSA is being legislated worldwide, and a number of standardization
bodies are working on it \cite{10.1109/MCOM.2013.6476873}.


The two most prominent types of DSA are the orthogonal spectrum
sharing (OSS) and the non-orthogonal spectrum sharing (NSS)
\cite{10.1109/MCOM.2014.6766097}.  In OSS, the operators coordinate
the shared bands (using, e.g., the X2 interface in LTE), so that a
given band is used \emph{exclusively by a single operator} at a given
time and place.  In NSS, a given band is used \emph{simultaneously by
a number of operators} at a given time and place.  In NSS, spectrum
sensing is key to learn of used and unused bands, and to minimize
radio interference.


PONs are successful mainly because of the cost-effective tree
topology.  First, the feeder fiber starts at the optical line terminal
(OLT) in the central office (CO), and ends at the first remote node
(RN) in some district.  From there, the distribution fibers lead to
further RNs in various neighborhoods, possibly through further RNs.
Finally, the last-mile fibers deliver the service to customer
premises.


NG-PONs should support direct communication between ONUs, without the
OLT relaying the data, in order to support direct communication
between BSs (connected to ONUs) required by future RANs.  However, in
legacy PONs, ONUs do not communicate directly with each other, but
through the OLT.  To this end, in \cite{10.1109/JLT.2010.2050861} the
authors propose two novel NG-PON architectures.  Interestingly, the
authors propose to cleverly use a circulator as a passive RN, which
would allow for some limited communication between BSs without the
OLT.  Another solution is to use the active RNs, which would also
enable NG-PONs to have larger splitting ratios and longer reach
\cite{10.1364/JOCN.2.000028}.


PONs are vulnerable to service disruption, because of the tree
architecture.  Failure of the OLT or the feeder fiber brings down the
entire PON.  Making a PON resilient is becoming more important, but
requires expensive redundant infrastructure, fibers and hardware.  In
\cite{10.1109/MCOM.2014.6736740} the authors review PON resiliency
mechanisms and propose their own mechanism for cost-effective
resiliency on request.


\section{Interoperator FMN sharing}
\label{sharing}


There are two operators, Operator 1 (O1) and Operator 2 (O2) who want
to share their FMNs, including spectrum.  An operator owns its FMN and
spectrum independently of the other operator.


In the proposed sharing we introduce the interoperator communication
(IC) to the access network.  The traffic of O1 accepted by the access
network of O2 is forwarded back to the network of O1 through the
aggregation network.  The IC makes the O1 service resilient to major
failures of its access network (like a power outage at a CO), which
otherwise would bring the service down.

\begin{figure}
  \centering
  \begin{tikzpicture}
    [node distance = 0.75 cm and 0.75 cm, font = \footnotesize]

    \node [mycloud] (internet) at (0, 0)
          {Internet};
    \node [router, font = \Large] (r1)
          [below left = of internet]
          {$\times$};
    \node [router, font = \Large, fill = black!30] (r2)
          [below right = of internet]
          {$\times$};

    \draw (r1) -- (internet.200);
    \draw [dashed] (r2) -- (internet.-20);

    \node [align = center] (r1i) [right = 0 cm of r1]
          {O1\\default\\router};
    \node [align = center] (r2i) [left = 0 cm of r2]
          {O2\\default\\router};

    \node [switch, font = \Large] (s1)
          [below = of r1]
          {$\times$};
    \node [switch, font = \Large, fill = black!30] (s2)
          [below = of r2]
          {$\times$};

    \node [align = right] (es) [left = 0 cm of s1]
          {Ethernet\\switch};

    \draw (r1) -- (s1);
    \draw [dashed] (r2) -- (s2);
    \draw [double] (s1) -- node [above] {interoperator trunk} (s2);

    \node [fmn] (fmn1) [below = of s1] {O1\\FMN};
    \node [fmn, fill = black!30] (fmn2) [below = of s2] {O2\\FMN};

    \draw (s1) -- (fmn1);
    \draw [dashed] (s2) -- (fmn2);

    \node [red] (ic) at ($ (fmn1)!0.5!(fmn2) $) {IC};
    \draw [icwave] (ic) -- (fmn1);
    \draw [icwave] (ic) -- (fmn2);

    \coordinate (lbr) at ($ (fmn1.west) - (1 cm, 0) $);
    \coordinate (rbr) at (fmn2.east);

    \coordinate (cla) at ($ (internet.north) + (0, 0.25 cm) $);
    \coordinate (cla1) at (cla -| lbr);
    \coordinate (cla2) at (cla -| rbr);
    \draw [dotted] (cla1) -- (cla2);

    \coordinate (clb) at ($ (r1.south)!0.5!(s1.north) $);
    \coordinate (clb1) at (clb -| lbr);
    \coordinate (clb2) at (clb -| rbr);
    \draw [dotted] (clb1) -- (clb2);

    \coordinate (clc) at ($ (s1.south)!0.5!(fmn1.north) $);
    \coordinate (clc1) at (clc -| lbr);
    \coordinate (clc2) at (clc -| rbr);
    \draw [dotted] (clc1) -- (clc2);

    \coordinate (cld) at ($ (fmn1.south) - (0, 0.25 cm) $);
    \coordinate (cld1) at (cld -| lbr);
    \coordinate (cld2) at (cld -| rbr);
    \draw [dotted] (cld1) -- (cld2);

    \draw [decorate, decoration = brace]
          (clb1) -- node [left = 10 pt, align = right]
          {IP\\network} (cla1);

    \draw [decorate, decoration = brace]
          (clc1) -- node [left = 10 pt, align = right]
          {aggregation\\network} (clb1);

    \draw [decorate, decoration = brace]
          (cld1) -- node [left = 10 pt, align = right]
          {access\\network} (clc1);

    \def\dfnca{0.125 cm}
    \def\dfncb{0.2 cm}

    \draw [dash pattern = on 0pt off 5pt on 9.225pt off 5pt, line width =
      3 pt, red, opacity = 0.5, rounded corners = 4 pt, line cap =
      round]
    ($ (fmn1.east) + (1.5 pt, \dfncb) $) --
    ($ (fmn2) + (-\dfnca, \dfncb) $) --
    ($ (s2) + (-\dfnca, -\dfnca - 0.5 pt) $) --
    ($ (s1) + (\dfnca, -\dfnca - 0.5 pt) $) --
    ($ (r1.south) + (\dfnca, -1.5 pt) $);

    \draw [dash pattern = on 0 off 4.9 pt, line width = 3 pt, green,
      opacity = 0.5, rounded corners = 4 pt, line cap = round]
    ($ (fmn1.north) + (-\dfnca, 1.5 pt) $) --
    ($ (r1.south) + (-\dfnca, -1.5 pt) $);

  \end{tikzpicture}
  \caption[General network architecture]{General network architecture.}
  \label{general}
\end{figure}
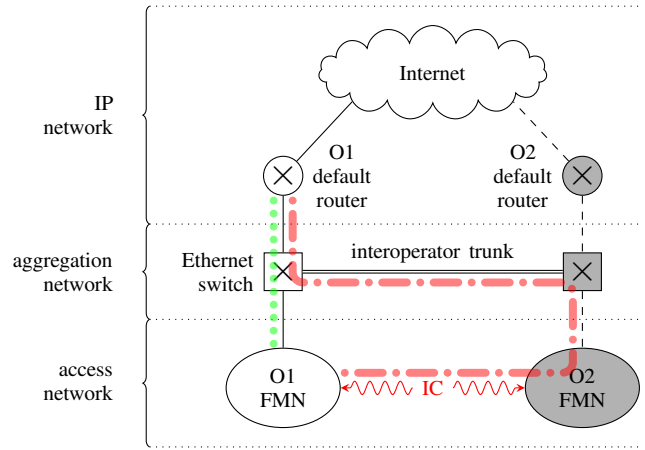
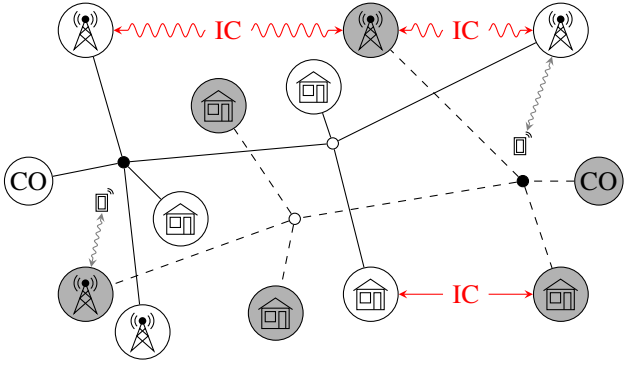
\begin{figure}
  \centering
  \begin{tikzpicture}[fill = black]

    \node [circ1] (co1) at (0, 0) {CO};
    \node [circ2] (co2) at (7.5, 0) {CO};

    \node [prn] (cs1a) at ($ (co1) + (1.25, 0.25) $) {};
    \node [arn] (cs1b) at ($ (co1) + (4, 0.5) $) {};
    \node [prn] (cs2a) at ($ (co2) - (1, 0) $) {};
    \node [arn] (cs2b) at ($ (co2) - (4, 0.5) $) {};

    \node [circ1] (bs1a) at (0.75, 2) {\BS{scale = 0.1}};
    \node [circ1] (bs1b) at (7, 2) {\BS{scale = 0.1}};
    \node [circ1] (bs1d) at (1.5, -2) {\BS{scale = 0.1}};
    \node [circ1] (ho1a) at (2, -0.5) {\FU{scale = 0.09}};
    \node [circ1] (ho1b) at (3.75, 1.25) {\FU{scale = 0.09}};
    \node [circ1] (ho1c) at (4.5, -1.5) {\FU{scale = 0.09}};

    \node [circ2] (bs2b) at (4.5, 2) {\BS{scale = 0.1}};
    \node [circ2] (bs2c) at (0.75, -1.5) {\BS{scale = 0.1}};
    \node [circ2] (ho2a) at (3.25, -1.75) {\FU{scale = 0.09}};
    \node [circ2] (ho2b) at (2.5, 1) {\FU{scale = 0.09}};
    \node [circ2] (ho2c) at (7, -1.5) {\FU{scale = 0.09}};

    \node [circle, inner sep = -0.2 mm] (mu1)
          at (6.5, 0.5) {\MU{scale = 0.15}};
    \node [circle, inner sep = -0.2 mm] (mu2)
          at (1, -0.25) {\MU{scale = 0.15}};
    
    \draw (co1) -- (cs1a) -- (cs1b);
    \draw (cs1a) -- (bs1a);
    \draw (cs1a) -- (bs1d);
    \draw (cs1a) -- (ho1a);
    \draw (cs1b) -- (bs1b);
    \draw (cs1b) -- (ho1b);
    \draw (cs1b) -- (ho1c);

    \draw[dashed] (co2) -- (cs2a) -- (cs2b);
    \draw [dashed] (cs2a) -- (ho2c);
    \draw [dashed] (cs2a) -- (bs2b);
    \draw [dashed] (cs2b) -- (ho2a);
    \draw [dashed] (cs2b) -- (bs2c);
    \draw [dashed] (cs2b) -- (ho2b);

    \node [red] (ic1) at ($ (bs1b)!0.5!(bs2b) $) {IC};
    \draw [icwave] (ic1) -- (bs1b);
    \draw [icwave] (ic1) -- (bs2b);
    \node [red] (ic2) at ($ (bs1a)!0.5!(bs2b) $) {IC};
    \draw [icwave] (ic2) -- (bs1a);
    \draw [icwave] (ic2) -- (bs2b);

    \node [red] (ic3) at ($ (ho1c)!0.5!(ho2c) $) {IC};
    \draw [iclink] (ic3) -- (ho1c);
    \draw [iclink] (ic3) -- (ho2c);

    \draw [muwave] (mu1) -- (bs1b);
    \draw [muwave] (mu2) -- (bs2c);

  \end{tikzpicture}
  \caption[Interoperator FMN sharing]{Interoperator FMN sharing, where
    CO is a central office, IC is the interoperator communication,
    $\bullet$ is a passive remote node, $\circ$ is an active remote
    node, \MU{scale = 0.12} is a mobile user, \BS{scale = 0.05} is a
    base station, and \FU{scale = 0.05} is a fixed user.}
  \label{access}
\end{figure}

Figures \ref{general} and \ref{access} show the network architecture
under study.  For O1, nodes are filled white and links are drawn
solid, and for O2, nodes are filled gray and links are drawn dashed.


The general network architecture is shown in Fig.~\ref{general}, where
O1 and O2 share their networks.  To keep the example simple, the
access network has only two FMNs, the aggregation network has only two
Ethernet switches, and the Internet Protocol (IP) network has only two
default routers.  The thick dotted path shows the working path of the
frames, and the thick dash-dotted path shows the backup path provided
by the IC.

At the IP layer, MUs and FUs of the FMNs communicate with the default
routers of their operators.  No IP routing is carried out in the
access or aggregation networks, since these networks switch Ethernet
frames only, which is a valid assumption for aggregation and access
networks.

In the FMNs, operators share their networks using the IC.  In the
aggregation network, operators share their networks using
interoperator Ethernet trunks, which carry modified Ethernet frames.
The Ethernet frames are modified, so that they can be sent over the
Ethernet network of the other operator.  Frames can be modified using
stacked VLANs (Q-in-Q) or stacked MACs (Mac-in-Mac); both should do
its job, but Mac-in-Mac would be better suited for large scale
deployments.


We abstract the details of specific technologies and make core
assumptions in order to take into account the currently deployed
networks (e.g., LTE, PONs) and the plausible future networks (e.g.,
LTE-A, NG-PON, RoF).  We describe the assumption about the RAN first,
and then about the backhaul.


As for the RANs, we assume that a BS is connected to the backhaul with
a single fiber.  A BS carries out the communication with MUs, and with
the BSs of the other operator.  The MU equipment is unaware of the IC
which is taking place between BSs, and so there is no need to modify
the MU equipment.  In the case of OSS, the BS software would have to
be upgraded to enable our proposed sharing, but without the need to
install new hardware, like the spectrum sensing hardware as would be
required in the case of NSS.


As for the backhaul, we assume there is one point of connection of the
backhaul to the central office (e.g., OLT in PONs).  We assume the
downstream and upstream throughput of the backhaul, e.g., 10 Gb/s or
100 Gb/s, is shared between a large number (e.g., 1024) of clients
(e.g., ONUs in PONs) which are either BSs or FUs.  We assume the tree
architecture of the optical distribution network (ODN).


We need active RNs to implement the proposed network sharing, because
they are able to accomplish what passive RNs cannot: diverging
upstream traffic to a detour downstream path if an upstream path
fails.  It is hard to argue for active RNs in PONs, because PONs are
successful mainly because of its passive ODN with passive RNs, which
are cheaper and more robust than active RNs.  Nevertheless, we rely on
active RNs, because we believe they will become more spread for two
important reasons.  First, active RNs can implement the direct
communication between ONUs required by 5G.  Second, active RNs (e.g.,
range extenders) are already used in PONs, and they are likely to be
more popular with NG-PONs, which can have a large splitting ratio
(e.g., 1:32) and be long-reach (above 100 km).

Fig.~\ref{access} shows the IC between two FMNs.  Each of the
operators has a CO at which an ODN of the tree topology is rooted.
FUs and BSs are connected to ODNs, and MUs communicate with BSs.  The
IC is taking place wirelessly between BSs of different operators and
optically between fixed users of different operators.


\section{Evaluation scenarios}
\label{scenarios}

We consider two evaluation scenarios.  They both are very similar and
differ only in the way the RN type is chosen.

\subsection{First scenario}

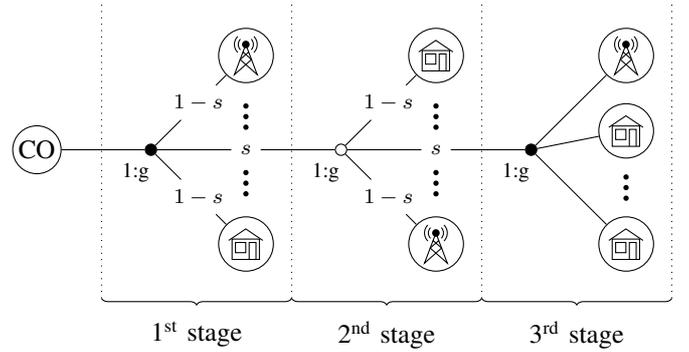
\begin{figure}
  \centering
  \begin{tikzpicture}[fill = black]

    \node [circ1] (co) at (0, 0) {CO};

    \node [prn] (rn1) at ($ (co) + (1.5, 0) $) {};
    \node [arn] (rn2) at ($ (co) + (4, 0) $) {};
    \node [prn] (rn3) at ($ (co) + (6.5, 0) $) {};

    \node [font = \footnotesize] at ($ (rn1) - (0.2 cm, 0.3 cm) $) {1:g};
    \node [font = \footnotesize] at ($ (rn2) - (0.2 cm, 0.3 cm) $) {1:g};
    \node [font = \footnotesize] at ($ (rn3) - (0.2 cm, 0.3 cm) $) {1:g};

    \def\d2r{1.25 cm}

    \node [circ1] (r1a) at ($ (rn1) + (\d2r, 1.25 cm) $) {\BS{scale = 0.1}};
    \node [circ1] (r1b) at ($ (rn1) + (\d2r, -1.25 cm) $) {\FU{scale = 0.09}};
    \node at ($ (r1a.south)!0.25!(r1b.north) $) {\VDOTS{}};
    \node at ($ (r1a.south)!0.75!(r1b.north) $) {\VDOTS{}};

    \node [circ1] (r2a) at ($ (rn2) + (\d2r, 1.25 cm) $) {\FU{scale = 0.09}};
    \node [circ1] (r2b) at ($ (rn2) + (\d2r, -1.25 cm) $) {\BS{scale = 0.1}};
    \node at ($ (r2a.south)!0.25!(r2b.north) $) {\VDOTS{}};
    \node at ($ (r2a.south)!0.75!(r2b.north) $) {\VDOTS{}};

    \node [circ1] (r3a) at ($ (rn3) + (\d2r, 1.25 cm) $) {\BS{scale = 0.1}};
    \node [circ1] (r3b) at ($ (rn3) + (\d2r, 0.25 cm) $) {\FU{scale = 0.09}};
    \node [circ1] (r3c) at ($ (rn3) + (\d2r, -1.25 cm) $) {\FU{scale = 0.09}};
    \node at ($ (r3b)!0.5!(r3c) $) {\VDOTS{}};

    \draw (co) -- (rn1);
    \node [font = \footnotesize] (s1a)
          at ($ (rn1)!0.5!(r1a) $) {$1 - s$};
    \draw (rn1) -- (s1a) -- (r1a);
    \node [font = \footnotesize] (s1b)
          at ($ (rn1)!0.5!(rn2) $) {$s$};
    \draw (rn1) -- (s1b) -- (rn2);
    \node [font = \footnotesize] (s1c)
          at ($ (rn1)!0.5!(r1b) $) {$1 - s$};
    \draw (rn1) -- (s1c) -- (r1b);

    \node [font = \footnotesize] (s2a)
          at ($ (rn2)!0.5!(r2a) $) {$1 - s$};
    \draw (rn2) -- (s2a) -- (r2a);
    \node [font = \footnotesize] (s2b)
          at ($ (rn2)!0.5!(rn3) $) {$s$};
    \draw (rn2) -- (s2b) -- (rn3);
    \node [font = \footnotesize] (s2c)
          at ($ (rn2)!0.5!(r2b) $) {$1 - s$};
    \draw (rn2) -- (s2c) -- (r2b);

    \draw (rn3) -- (r3a);
    \draw (rn3) -- (r3b);
    \draw (rn3) -- (r3c);

    \coordinate (d1) at ($ (rn1) - (0.65 cm, 1.97 cm) $);
    \coordinate (d2) at ($ (rn2) - (0.65 cm, 1.97 cm) $);
    \coordinate (d3) at ($ (rn3) - (0.65 cm, 1.97 cm) $);
    \coordinate (d4) at ($ (rn3) + (1.85 cm, -1.97 cm) $);
    \draw [dotted] (d1) -++ (0, 3.94 cm);
    \draw [dotted] (d2) -++ (0, 3.94 cm);
    \draw [dotted] (d3) -++ (0, 3.94 cm);
    \draw [dotted] (d4) -++ (0, 3.94 cm);
    \draw [decorate, decoration = brace]
          (d2) -- node [below = 5 pt, align = center]
          {\nth{1} stage} (d1);
    \draw [decorate, decoration = brace]
          (d3) -- node [below = 5 pt, align = center]
          {\nth{2} stage} (d2);
    \draw [decorate, decoration = brace]
          (d4) -- node [below = 5 pt, align = center]
          {\nth{3} stage} (d3);

  \end{tikzpicture}
  \caption[PON topology model]{PON topology model, where CO is a
    central office, $\bullet$ is a passive remote node, $\circ$ is an
    active remote node, \BS{scale = 0.05} is a base station, and
    \FU{scale = 0.05} is a fixed user.}
  \label{topology}
\end{figure}

Fig.~\ref{topology} illustrates the first scenario.  The PON has the
depth of three stages.  The PON can have many second and third stages,
but in the figure we show only one of each.  In the first stage a
passive RN with the \mbox{1:g} splitting ratio is used.  The
probability that a fiber coming out of the RN goes to a second stage
is $s$ and, conversely, $(1 - s)$ that it goes to an ONU, a FU or a
BS.

The high \mbox{1:g} splitting ratio and possibly long feeder and
distribution fibers may require an active RN, and so at the second
stage we install an active RN.  At the second stage the probabilities
of $s$ and $(1 - s)$ have the same meanings.  Finally, at the third
stage there is a passive RN installed, and all fibers reach an ONU.
The last-mile fibers are typically short, and even with a \mbox{1:g}
or higher splitting ratio, passive RNs suffice.

An ONU is capable of the IC with probability $r$.  These
interoperator-communicating ONUs (IC-ONUs) can offer the Internet
communication in the same way as the OLT does, while the remaining
ONUs are non interoperator-communicating ONUs (NIC-ONUs).

For this PON topology of depth three, the given $s$, and the given
\mbox{1:g} splitting ratio, the mean number $N$ of ONUs is given by
(\ref{e:N}).

\begin{equation}
  N = g(1 - s + gs(1 - s + gs))
  \label{e:N}
\end{equation}

We assume $s = 0.3$ and $g = 32$, and so the number of ONUs is $N
\approx 3187$, which is reasonable for NG-PONs.  For instance,
currently the XG-PON supports 1024 ONUs.

The availability values of the PON components are taken from
\cite{10.1109/MCOM.2014.6736740}, and they are summarized in Table
\ref{assumptions}.  The reported availability value for the passive RN
is that for the \mbox{1:32} power splitter, and for the active RN is
that for the OLT.  The values reported for the feeder fiber, the
distribution fiber and the last-mile fiber are calculated with the
reported fiber availability per km assuming that their mean lengths
are 10 km, 3 km and 0.7 km, respectively.

\begin{table}
  \centering
  \caption{Availability values.}
  \label{assumptions}
  \begin{tabular}{|r|l|}
    \hline
    \multicolumn{1}{|c|}{\bf{}Component} & {\bf{}Availability} \\
    \hline
    OLT & 0.9999485 \\
    ONU & 0.9999645 \\
    passive remote node & 0.9999987 \\
    active remote node & 0.9999485 \\
    fiber per km & 0.9999429 \\
    feeder fiber & 0.999429 \\
    distribution fiber & 0.999829 \\
    last-mile fiber & 0.99996 \\
    \hline
  \end{tabular}
\end{table}

\subsection{Second scenario}

The second scenario differs from the first scenario only in the
selection of the RN types, i.e., whether they are passive or active.
While in the first scenario the type of a RN is given up front, in the
second scenario it is given probabilistically: a RN is active with
probability $q$.  The second scenario allows us to study how the ONU
service availability changes as a function of probability $q$.


\section{Service availability calculation}
\label{evaluation}


We want to calculate the mean ONU service availability (SA) for a
given network taken from the network populations of the two scenarios.
For the given network, we are provided the topology, the type of RNs,
and the information on which ONUs are capable of the IC.  The mean ONU
SA is the arithmetic mean of the SAs of all ONUs (all IC-ONUs and
NIC-ONUs).  The SA of the IC-ONU equals the availability of the OLT,
because it can rely on the IC.  The problem is to calculate the SA of
a given NIC-ONU.


The SA calculation for the proposed sharing is more difficult than for
traditional PONs.  The ONU SA for traditional PONs is calculated by
following upstream a single path from the ONU to the OLT, and just
multiplying the availabilities of the encountered components.  In the
proposed sharing, the availability calculation is more complicated for
three reasons.  First, in addition to the path from the NIC-ONU to the
OLT, we need to consider the paths from the NIC-ONU to all the
IC-ONUs.  Second, the considered paths are not always upstream only: a
path can be upstream-downstream at the same time, i.e., it can go
upstream first and downstream next to reach an IC-ONU.  Third, an
upstream-downstream path can traverse some nodes and fibers twice, and
their availabilities should be taken into account only once.


The SA of the given NIC-ONU is calculated by evaluating a reliability
block diagram (RBD) of the service paths from the NIC-ONU to the OLT
and all the IC-ONUs.  Since a PON has the tree topology, the
corresponding RBDs have the parallel and serial configurations only,
without the crossover configurations, making the evaluation easy to
implement programmatically with the recursive depth-first search.
Nonetheless, the evaluation has some important intricacies, and we
discuss them further below.


The recursive function $f(c, p)$ calculates the SA for the current
node $c$, provided the previous node is $p$.  Node $p$ preceded the
current node $c$, i.e., node $p$ was the current node in the previous
call of the function.  The function is initially called with the
NIC-ONU of interest as the current node, and with $p = \text{null}$.
The function recursively calls itself to calculate the availabilities
of the RNs, and eventually of the OLT and other ONUs.

Function $f(c, p)$ is given by (\ref{f}), where $a_c$ is the
availability of node $c$, $u_c$ is the node upstream of node $c$, $u_c
\to c$ is the upstream fiber of node $c$, $a_{u_c \to c}$ is the
availability of that fiber, and $N_c$ is the set of neighbor nodes of
node $c$.  Symbols $V_c$, $h_c$, and $d_{c, v}$ are defined further
down.

\begin{equation}
  f(c, p) = \left\{
  \begin{array}{cr}
    a_ca_{u_c \to c}f(u_c, c) & \text{\nth{1} case}\\[15pt]
    0 & \text{\nth{2} case}\\[15pt]
    a_c & \text{\nth{3} case}\\[15pt]
    a_c(1 - \prod\limits_{\substack{i \in N_c\\i \ne p}}(1 - a_{i \to c}f(i, c))) &
    \text{\nth{4} case}\\[15pt]
    h_c(1 - \prod\limits_{v \in V_c}(1 - d_{c, v})) &
    \text{\nth{5} case}
  \end{array}\right.
  \label{f}
\end{equation}

The cases of the function are as follows:

\begin{itemize}

\item[\nth{1}] case is for the initial call of the function, i.e.,
  when $c$ is an NIC-ONU and $p = \text{null}$, which allows the
  function to reach the upstream node $u_c$,

\item[\nth{2}] case applies when the function reaches an NIC-ONU from
  some previous node, i.e., $p \ne \text{null}$, in which case no
  service is offered,

\item[\nth{3}] case applies when $c$ is the OLT or an IC-ONU, which
  offer the service,

\item[\nth{4}] case applies when $c$ is an active RN or a passive RN
  reached from an upstream node, i.e., $p = u_c$, which offers to
  reach in parallel the neighbor nodes $N_c$ of node $c$, excluding
  node $p$,

\item[\nth{5}] case applies when $c$ is a passive RN reached from a
  downstream node, i.e., $p \ne u_c$, which is the most difficult
  case discussed below.

\end{itemize}


In the \nth{5} case the upstream-downstream paths exist for an NIC-ONU
connected to a passive RN, when to the same passive segment (i.e., a
sequence of passive RNs) there are IC-ONUs connected.  The NIC-ONU can
get service either from nodes reachable through the first active
upstream node (the OLT or an active RN), or from an IC-ONU connected
to the same passive segment.  All paths for the NIC-ONU have the same
\emph{shared path}, starting at the first passive upstream RN for
which the function was called, through the upstream fibers and
possibly further passive upstream RNs, up to and including the first
upstream active node.  The availability of the shared path for node
$c$ is $h_c$, and it has to be accounted for only once.  From this
shared path all \emph{parallel non-shared paths} $V_c$ fork, i.e., the
paths for service nodes reachable from the first active node, and the
paths for the IC-ONUs connected to the same passive segment.  The
availability of the parallel non-shared path $v$ for node $c$ is
$d_{c, v}$.

\begin{figure}
  \centering
  \begin{tikzpicture}[fill = black]

    \node [circle] (co) at (0, 0) {\VDOTS{rotate = 90}};

    \node [arn] (rn1) at ($ (co) + (0.9, 0) $) {};
    \node [prn] (rn2) at ($ (co) + (2.7, 0) $) {};
    \node [prn] (rn3) at ($ (co) + (4.5, 0) $) {};
    \node [prn] (rn4) at ($ (co) + (6.3, 0) $) {};

    \node [font = \footnotesize] at ($ (rn1) - (0.2 cm, 0.3 cm) $) {RN1};
    \node [font = \footnotesize] at ($ (rn2) - (0.2 cm, 0.3 cm) $) {RN2};
    \node [font = \footnotesize] at ($ (rn3) - (0.2 cm, 0.3 cm) $) {RN3};
    \node [font = \footnotesize] at ($ (rn4) - (0.2 cm, 0.3 cm) $) {RN4};

    \def\d2r{1 cm}

    \draw (co) -- (rn1);
    \node [circle] (r1a) at ($ (rn1) + (0.75 cm, 0.75 cm) $) {\VDOTS{rotate = -45}};
    \node [circle] (r1b) at ($ (rn1) + (0.75 cm, -0.75 cm) $) {\VDOTS{rotate = 45}};
    \draw (rn1) -- (r1a);
    \draw (rn1) -- (rn2);
    \draw (rn1) -- (r1b);

    \node [circ1, label = IC-ONU1] (r2a) at ($ (rn2) + (\d2r, \d2r) $) {\BS{scale = 0.1}};
    \node [circle] (r2b) at ($ (rn2) + (0.75 cm, -0.75 cm) $) {\VDOTS{rotate = 45}};
    \draw (rn2) -- (r2a);
    \draw (rn2) -- (rn3);
    \draw (rn2) -- (r2b);

    \node [circ1, label = NIC-ONU1] (r3a) at ($ (rn3) + (\d2r, \d2r) $) {\FU{scale = 0.09}};
    \node [circle] (r3b) at ($ (rn3) + (0.75 cm, -0.75 cm) $) {\VDOTS{rotate = 45}};
    \draw (rn3) -- (r3a);
    \draw (rn3) -- (rn4);
    \draw (rn3) -- (r3b);

    \node [circ1, label = IC-ONU2] (r4a) at ($ (rn4) + (\d2r, \d2r) $) {\BS{scale = 0.1}}; 
    \node [circle] (r4b) at ($ (rn4) + (0.75 cm, -0.75 cm) $) {\VDOTS{rotate = 45}};
    \draw (rn4) -- (r4a);
    \draw (rn4) -- (r4b);

    \draw [line width = 10 pt, green, opacity = 0.5, line cap = round]
    ($ (rn1.west) + (1.5 pt, 0) $) -- ($ (rn3.east) - (1.5 pt, 0) $);

    \draw [dashed] plot [smooth cycle, tension = 0.65] coordinates {
      (0.5, 0)
      (0.7, 0.3) (1.7, 0.4) (2.1, 0.8) (1.75, 1.2) (0.8, 1.15)
      (-0.25, 0.5) (-0.25, -0.5)
      (0.8, -1.15) (1.75, -1.2) (2.1, -0.8) (1.7, -0.4) (0.7, -0.3)};

    \draw [dashed] plot [smooth cycle, tension = 0.65] coordinates {
      ($ (rn2) + (0.25, 0.25) $)
      ($ (r2a) + (0.3, -0.3) $)
      ($ (r2a) + (0.3, 0.3) $)
      ($ (r2a) + (-0.3, 0.3) $)};

    \draw [dashed] plot [smooth cycle, tension = 0.65] coordinates {
      ($ (r4a) + (0.3, 0.3) $)
      ($ (r4a) + (-0.3, 0.3) $)
      ($ (rn4) + (0, 0.3) $)
      ($ (rn3) + (0.5, 0.13) $)
      ($ (rn3) + (0.25, 0) $)
      ($ (rn3) + (0.5, -0.13) $)
      ($ (rn4) + (-0.3, -0.15) $)
      ($ (rn4) + (0.4, -0.025) $)
      ($ (r4a) + (0.3, -0.3) $)};

  \end{tikzpicture}
  \caption[Example for the \nth{5} case]{Example for the \nth{5} case
    of function $f(c, p)$, where the thick line segment shows the
    shared path, the dashed curves show the non-shared paths,
    $\bullet$ is a passive remote node, and $\circ$ is an active
    remote node.}
  \label{5th}
\end{figure}
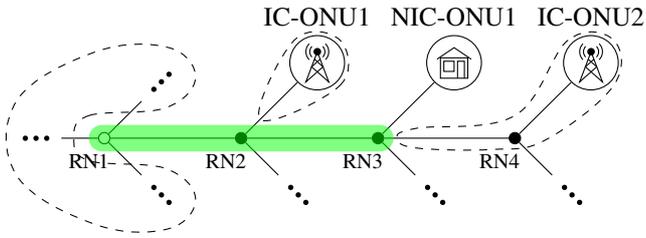

Fig.~\ref{5th} shows an example for the \nth{5} case of function $f(c,
p)$.  We calculate the SA of node NIC-ONU1, which can get service from
nodes reachable through node RN1, or from nodes IC-ONU1 and IC-ONU2.
The function is called for node NIC-ONU1 first, and for node RN3 next,
i.e., $c = \text{RN3}$.  The thick line segment highlights the shared
path with availability $h_c$, and the dashed curves show the parallel
non-shared paths $V_c$ with availabilities $d_{c, v}$.


\section{Evaluation results}
\label{results}

We evaluated the ONU SA by randomly generating network samples for the
network populations with the given characteristics for the two
scenarios.  Each network sample has one hundred networks.  The sample
mean of the ONU SA is the arithmetic mean of the ONU SAs calculated
for the networks in the sample.  We deem the sample means credibly
estimate the population means since the relative standard errors of
all sample means are below 1\%.  For both scenarios there were 874
populations considered, and 87400 networks evaluated.  The software is
available at \cite{availawebsite}.

\subsection{First scenario results}

For the first scenario the varying characteristic of populations is
probability $r = 0, 10^{-3}, 2 \cdot 10^{-3}, \ldots, 10^{-2}, 2 \cdot
10^{-2}, \ldots, 10^{-1}, 1.5 \cdot 10^{-1}, \ldots, 1$ with 38
values.  Besides the first scenario populations, we consider also the
traditional PON populations (i.e., all RNs are passive), and allow for
the IC with probability $r$.  And so we consider $2 \cdot 38 = 76$
populations, and evaluate 7600 networks.

\begin{figure}
  \begin{tikzpicture}
    \begin{semilogxaxis}
      [width = 0.9\columnwidth, height = 6 cm,
        xlabel = $r$, ylabel = availability,
        legend style = {font = \small, inner sep = 2 pt, text height = 0.9 ex},
        legend pos = north west, legend columns = 2,
        y label style = {at = {(axis description cs:-0.045,.5)}},
        ytick = {0.999, 0.9995, 1}, yticklabels = {0.999, 0.9995, 1}]

      \addplot [solid, mark = *, mark size = 0.75 pt] table {mix.txt};

      \addplot [dashdotted, no markers] coordinates {(0.001, 0.998921) (1, 0.998921)};

      \addplot [dotted, mark = *, mark size = 0.75 pt, mark options = {solid}] table {prn.txt};

      \addplot [dashed, no markers] coordinates {(0.001, 0.99897) (1, 0.99897)};

      \legend{\nth{1} scenario, $r = 0$, traditional, baseline}
    \end{semilogxaxis}
  \end{tikzpicture}
  \caption{Availability in the \nth{1} scenario as a function of $r$.}
  \label{fsa}
\end{figure}
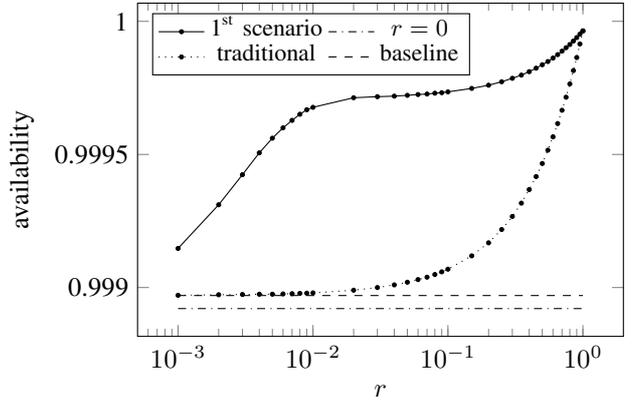

Fig.~\ref{fsa} shows the SA results in a logarithmic scale as a
function of probability $r$.  Each data point represents the sample
mean.  The line marked as ``$r = 0$'' shows the SA of 0.998921 for the
first scenario with $r = 0$, and the ``baseline'' shows the SA of
0.99897 for a traditional PON.  Without the IC, i.e., $r = 0$, the
introduction of the active RNs decreases the SA, because the
availability of an active RN is smaller than that of a passive RN.

However, even a small increase in $r$ causes a large increase in the
SA for the first scenario, but a little increase for the traditional
PON with the IC.  For instance, when $r = 10^{-2}$, i.e., when on
average one ONU out of a hundred has the IC, the SA is 0.999677, which
corresponds to 2.83 hours of downtime per year, while the baseline SA
of the traditional PON is 0.99897, which corresponds to 9.02 hours of
downtime per year.  This is more than a threefold reduction in
downtime.

For the first scenario networks, the SA rapidly increases as $r$
increases, because the presence of a single IC-ONU in a passive
segment of the third stage allows all NIC-ONUs in that segment to
reach the IC-ONU through an active RN in the second stage.  This rapid
trend continues up to the point when most passive segments in the
third stage have an IC-ONU, and after that point adding more IC-ONUs
does not increase rapidly the SA, as shown by the nearly-flat SA for
$10^{-2} < r < 10^{-1}$.  For $r > 10^{-1}$ the SA increases linearly
because the high availability of the IC-ONUs increases directly the
mean ONU SA.

In the evaluation we allowed the traditional PONs to be capable of the
IC in order to show that the active RNs used in the first scenario
help realize the full potential of the IC, which is evident for the
small and practical values of $r$.  Without the active RNs, the SA
increases linearly as a function of $r$, as is the case for the
traditional PONs with the IC.

\subsection{Second scenario results}

For the second scenario the probability $r$ and probability $q$ vary.
Probability $r$ varies as in the first scenario, and $q = {0, 0.05,
0.1, \dots, 1}$ has 21 values.  Therefore there are $38 \cdot 21 =
798$ network populations considered, and 79800 networks evaluated.

\begin{figure}
  \begin{tikzpicture}
    \begin{semilogxaxis}
      [width = 0.9\columnwidth, height = 6.5 cm,
        view/h = -20, view/v = 10,
        xlabel = $r$, ylabel = $q$, zlabel = availability,
        x tick label style = {xshift = 2 pt, yshift = -1 pt},
        y tick label style = {xshift = -3 pt, yshift = 3 pt},
        ztick = {0.999, 0.9995, 1}, zticklabels = {0.999, 0.9995, 1}]

    \addplot3 [surf, faceted color = gray, fill = gray, opacity = 0.40] coordinates {
      (0.001, 0.001, 0.99897)
      (0.001, 1, 0.99897)

      (1, 0.001, 0.99897)
      (1, 1, 0.99897)};

    \addplot3 [mesh, draw = black] table {data.txt};

    \end{semilogxaxis}
  \end{tikzpicture}
  \caption{Availability in the \nth{2} scenario as a function of $r$
    and $q$.}
  \label{ssa}
\end{figure}
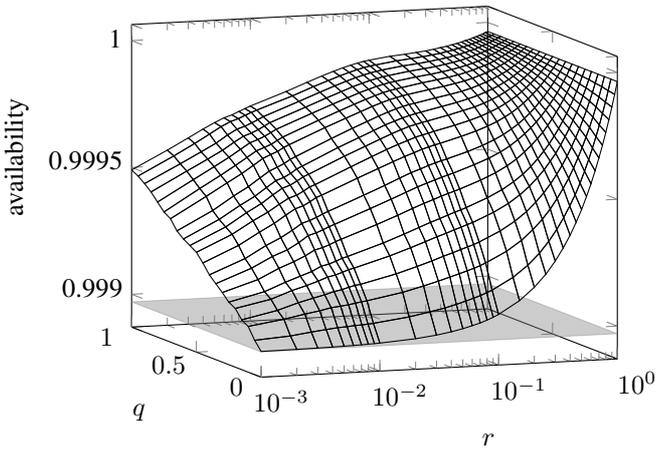

\begin{figure}
  \begin{tikzpicture}
    \begin{axis}
      [width = 0.9\columnwidth, height = 6 cm,
        xlabel = $q$, ylabel = availability,
        y label style = {at = {(axis description cs:-0.045,.5)}},
        ytick = {0.9988, 0.999}, yticklabels = {0.9988, 0.999},
        ymin = 0.99879, ymax = 0.99901, 
        legend style = {font = \small, inner sep = 2 pt, text height = 0.9 ex},
        legend pos = south west]

    \addplot [solid, mark = *, mark size = 0.75 pt] table {aw0r.txt};

    \addplot [dashed, no markers] coordinates {(0.001, 0.998971) (1, 0.998971)};

    \legend{$r = 0$, baseline}
    \end{axis}
  \end{tikzpicture}
  \caption{Availability in the \nth{2} scenario as a function of $q$
    with $r = 0$.}
  \label{aw0r}
\end{figure}
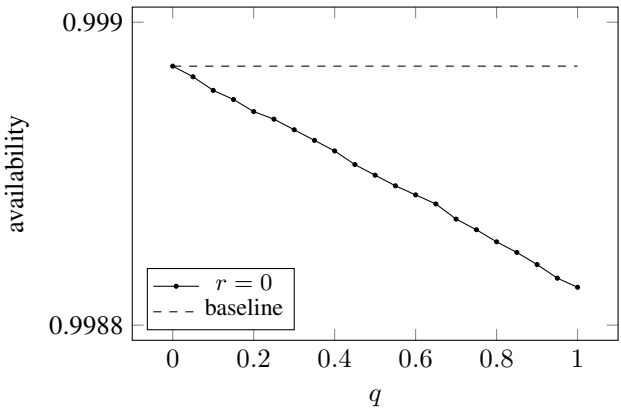

The results are shown in Fig.~\ref{ssa} for $r \ne 0$ because of the
logarithmic scale, and in Fig.~\ref{aw0r} for $r = 0$.  A data point
in the figures represents the sample mean of the SA.  The baseline SA
of 0.99897 of a traditional PON without the IC is shown in
Fig.~\ref{ssa} as the gray plane, and in Fig.~\ref{aw0r} as the dashed
line.

For $q = 0$, the SA is the same as for the traditional PONs in
Fig.~\ref{fsa}, because these networks do not have active RNs.  The SA
increases together with the increasing $r$ and $q$, but the increase
is not as impressive as for the first scenario, since in the first
scenario active RNs are positioned strategically in the second stage
where they can interconnect a large number of NIC-ONUs to a single
IC-ONU, while in the second scenario an active RN can land in the
third stage where it is less useful.

Fig.~\ref{aw0r} shows that the SA decreases when active RNs are
introduced, and when there are no IC-ONUs in the network, i.e., $r =
0$.  These results validate the SA calculation.

\section{Conclusion}
\label{conclusion}


We proposed the novel idea of interoperator fixed-mobile network
sharing, and showed that the service downtime can be reduced threefold
by introducing a small number of active remote nodes and providing
interoperator communication to as little as 1\% of all optical network
units.  The interoperator communication can be delivered wirelessly
between mobile base stations or with a fiber connecting optical
network units.


The proposed solution would require the installation of a few active
nodes and updating the software in network nodes, while the hardware
of regular users would require no changes.  The deployment of the
active nodes could be rolled out in stages when needed, and in those
areas where the resiliency is needed most, i.e., in a business
district.


The proposed sharing allows for dynamic reconfiguration, since the
interoperator communication can be carried out wirelessly by base
stations.  An operator can easily start or end the sharing with
various operators, which would encourage competition.


Future work could concentrate on
\begin{inparaenum}
\item generalizing the proposed sharing to other mobile backhaul
  types, like the microwave backhaul;
\item optimizing the placement of active remote nodes;
\item optimizing the selection of optical network units for
  interoperator communication;
\item optimizing various economic metrics, such as revenue, or the
  risk of liability due to service failure;
\item studying incentives which would foster sharing, and rules which
  would discourage cheating;
\item researching various aspects related to fixed-mobile networks,
  such as cognitive radio, various static and dynamic traffic models,
  cognitive radio, or coordinated multipoint transmission; and
\item generalizing the proposed sharing to any number of operators.
\end{inparaenum}

\section*{Acknowledgments}

This work was supported by the postdoctoral fellowship number
DEC-2013/08/S/ST7/00576 from the Polish National Science Centre.  The
numerical results were obtained using PL-Grid, the Polish
supercomputing infrastructure.

\bibliographystyle{IEEEtran}
\bibliography{all}

\end{document}